\documentstyle[12pt]{article}
\def\dj{d\kern-.30em\raise1.25ex\vbox{\hrule width .3em height .03em}}
\def\Dj{D\rlap{\kern-.70em\raise0.75ex\vbox{\hrule width .3em height .03em}}}

\begin{document}

\title{Generalized Noiseless Quantum Codes utilizing Quantum Enveloping Algebras}
\author{Micho {\Dj}ur{\dj}evich \\Instituto de Matematicas,
 UNAM, Mexico DF, 
CP 04510, Mexico\\
\and Hanna E. Makaruk \hspace*{2cm} Robert Owczarek\\  Los Alamos
National Laboratory, Los Alamos, NM 87545, USA}
\date{ }
\maketitle
\begin{abstract}
A generalization of the results of Rasetti and Zanardi concerning
avoiding errors in quantum computers by using states preserved by evolution
is presented. The concept of  dynamical symmetry is generalized from the level of
classical Lie algebras and groups to
the level of a dynamical symmetry based on quantum Lie algebras and quantum groups
(in the sense of Woronowicz).
 A natural connection is proved between states 
preserved
by representations of a quantum  group and states preserved by evolution with dynamical
symmetry of the appropriate universal enveloping algebra. 
Illustrative examples are discussed. 
\end{abstract}
The dynamical  symmetry of a system is  a valuable property.
One can apply to
systems  which possess a dynamical symmetry powerful methods 
based on  the theory of Lie algebras and
their representations, like the method of coherent states 
\cite{Pierelomov}.
Dynamical symmetry has also proved to be  important  in searching 
for physical systems with very specific quantum states: states which can not 
be corrupted by their interactions with the 
environment \cite{Rasetti}. These states can be used to provide noiseless quantum  codes that have great  potential 
utility  for constructing quantum computers.  Noiseless 
quantum   codes can be an alternative or a supplement to
 error correcting codes, which are elaborate methods for 
 coding information, recognizing 
errors and correcting them \cite{Shor,Steane,Laflamme}.

This paper  introduces the notion of a dynamical symmetry 
associated with quantum groups. We apply this concept of dynamical symmetry
to the study of systems which  demonstrate the usefulness of noiseless quantum codes. 

We first  outline the basic mathematical concepts and tools that 
will be used in the paper. Quantum groups will be understood as $C$*-Hopf algebras 
following \cite{Woronowicz1, Woronowicz2}.  However, in our further considerations only
 the  *-Hopf algebra structure of quantum groups is used;  
the $C$*-algebra structure is not important  in our applications. 

A *-Hopf algebra is  a complex unital
algebra $\cal{A}$ (with the unit $1_{\cal A}\in {\cal A}$)  
equipped with linear maps of the coproduct
$\Phi : {\cal A}\to{\cal A}\otimes {\cal A}$, counit
$e:{\cal A}\to {C\hspace{-2.8mm}I}$ and the antipode
 $\kappa :{\cal A}\to{\cal A}$, for which the following 
identities hold:
\begin{eqnarray}
(\mbox{\rm id}_{\cal A} \otimes \Phi)\Phi&=&(\Phi \otimes \mbox{\rm id}_{\cal A})
\Phi\nonumber\\
(e \otimes \mbox{\rm id}_{\cal A})\Phi(a)&=&( \mbox{\rm id}_{\cal A} \otimes e)\Phi(a)=a\nonumber\\
m(\kappa\otimes \mbox{\rm id}_{\cal A})\Phi(a)&=&m( \mbox{\rm id}_{\cal A} \otimes
 \kappa)\Phi(a)=e(a)1_{\cal A}\label{1}
\end{eqnarray}
where $m\colon\cal{A}\otimes\cal{A}\rightarrow\cal{A}$ 
is the product in $A$, explicitly given by $m(a\otimes b)=a\cdot b$. 
We also assume that there is an
antilinear involutive map $^*\colon{\cal A}\to{\cal A}$ 
which is antimultiplicative in the sense that $(a  b)^*= b^* a^*$,
 and which is compatible with 
$\Phi$ in the sense that  $\Phi ^*=(^*\otimes ^*)\Phi$. 

The whole classical theory of Lie groups and Lie algebras
can be viewed as a special case of the theory of quantum groups.
For classical groups, 
the algebra $\cal{A}$ will be commutative and consist of 
complex-valued
polynomial functions over the group. Furthermore,  in the
classical case
$^*$ is the standard complex conjugation of functions,
and the maps $\Phi$, $e$ and $\kappa$ represent the product
in the group, the neutral element,  and the operation
of taking inverses of the elements of the group, 
respectively. 
The algebraic identities (\ref{1}) correspond 
in the classical case 
to the associativity  of the group product 
(coassociativity of $\Phi$), multiplication of
 the element with its inverse giving the neutral 
element (\ref{1}), and multiplication of an element 
with the unit element giving original element
\cite{Woronowicz2}. General noncommutative $^*$-Hopf
algebras $\cal A$ have no such interpretation, but are considered
informally as  algebras of `functions' on more general
objects called {\it quantum groups} $G$. 

A generalization of the concept of dynamical symmetry can
be defined only when there are well-established notions 
of a Lie algebra and a corresponding universal enveloping algebra 
associated with each quantum group $G$,  including the 
corresponding 
representation theory. In quantum theory, all these notions 
 depend essentially on an
appropriately chosen {\it differential calculus} over $G$. 
 
First-order differential calculi are defined as certain {\it modules} $\Gamma$ 
over ${\cal A}$, equipped with 
a {\it differential} $d\colon{\cal A}\rightarrow \Gamma$. 
The module $\Gamma$ is a noncommutative counterpart of the 
usual module of $1$-forms over a classical group, 
and $d$ generalizes the standard differential of functions.

In  quantum group theory,  a special role is played by so-called {\it left-covariant} and
{\it bicovariant} differential calculi. In these cases,  we can introduce the analogs of left and
left/right actions of the group $G$ on the module \cite{Woronowicz3}. 
If the module $\Gamma$ is left-covariant, then we can define a subspace 
$\Gamma_{\mbox{\scriptsize\rm inv}}$ of $\Gamma$, which consists of
left-invariant `$1$-forms'. A quantum Lie algebra can  then be  defined as the corresponding
dual space, $L=\Gamma^*_{\mbox{\scriptsize\rm inv}}$. 

If the calculus is bicovariant, then we can introduce a natural {\it braid operator}
$\sigma\colon L\otimes L\rightarrow L\otimes L$, playing the role of the classical
transposition. Furthermore, 
in analogy with classical theory, we can 
define the Lie bracket in the space $L$ \cite{Woronowicz3}. 
The Lie bracket is defined by  an operator 
$C: L\otimes L\to L$, as
 $[\cdot, \cdot]:L\times L\to L$, 
$[x,y]=C(x\otimes y)$, and it satisfies an  appropriately  
generalized Jacobi identity.

Following  classical theory, 
the quantum universal enveloping algebra 
for $(L,[,])$ is defined as a unital associative 
algebra $U(L)$  generated by the relations 
$xy-\sum_iy_ix_i=[x,y],$
where $x,y\in L$ and $\sum_iy_i\otimes x_i=\sigma(x\otimes y)$. 

Having this bracket and using the above equation one can define
representations of quantum  Lie algebras and of the 
corresponding quantum  universal enveloping algebras.
 It can be shown that every representation $v$ of $G$ 
in a  finite-dimensional vector space $V$ naturally 
gives rise to a representation $S:U(L)\to\mbox{\rm End}(V)$ of  the quantum universal 
enveloping algebra.  
Namely,  let $v:V\to V\otimes {\cal A}$ be a (left)
 representation of the quantum group 
($^*$-Hopf algebra) ${\cal A}$ in a  finite dimensional 
complex vector space V (i.e. $v$ is 
linear and satisfies the conditions 
$(\mbox{\rm id}_V\otimes\Phi)v=(v\otimes \mbox{\rm id}_{\cal A})v$,
 and $(\mbox{\rm id}_V\otimes e)v=\mbox{\rm id}_V$,
 corresponding to the usual  requirements for representations
  of groups ---
 the products of group elements are represented by
   compositions  of operators representing these 
elements, and the neutral element of a group is 
represented by the identity operator). 
Every such representation of $G$ in $V$ 
naturally generates a representation 
$\delta: U(L)\to \mbox {\rm End}(V)$ of $U(L)$ in $V$ 
(if the differential  calculus is bicovariant)
 or only of the Lie algebra $L$, $\delta: L\to \mbox {\rm End}(V)$
 (if the differential calculus is left-covariant).

 Moreover,   if the differential calculus is $^*$-covariant,
 which  means that in the module $\Gamma$ of $1$-forms 
is defined the $^*$ -operation  $^*:\Gamma\to\Gamma$ 
induced by $^*$ in ${\cal A}$, it makes sense to 
speak about the hermiticity of the representation $\delta$.  
Namely, the $^*$-operation  on $\Gamma$ naturally 
induces the $^*$-structure on the quantum Lie 
algebra $L$, via the formula $<f^*,\psi>=-<f,\psi^*>$ 
where $f\in L=\Gamma^*_{\mbox{\scriptsize \rm inv}}$ 
and $\psi\in \Gamma^*_{\mbox{\scriptsize \rm inv}}$. 
 Moreover, if the quantum group is ``connected" in the
  sense  that 
${\rm ker}(d)=C\hspace{-2.8mm}I\cdot 1_{\cal A}$ , then
 one can prove that these two conditions are equivalent.

Suppose then, that 
 a $^*$-covariant, left-covariant 
differential calculus is defined on a quantum  group  $G$ and $L,U(L),V,v,\delta$
 are as above. We define an open system with
 quantum  dynamical symmetry as a system whose 
evolution is defined  in the Hilbert space $V$. 
The system interacts with its environment 
described by  the Hilbert space, $H_B$ 
(assumed for simplicity to be  finite-dimensional; however,  everything could 
be incorporated into the
infinite-dimensional case).    

We say that a system has quantum  dynamical 
symmetry  described by the quantum  group $G$ 
and its quantum  Lie algebra $L$ 
if the following conditions are 
 satisfied:

\begin{description}
\item{i)} The evolution of the system is 
governed by the Hamiltonian 
$$h\in \mbox{\rm End}(V\otimes H_B)\simeq 
\mbox{\rm End}(V)\otimes \mbox{\rm End}(H_B).$$
\item{ii)} The Hamiltonian is Hermitian $(h^*=h)$
 with $^*$ defined as the  tensor product of natural star operators described  above
acting 
in $\mbox{\rm End}(V)$ and $\mbox{\rm End}(H_B)$.
\item{iii)} The Hamiltonian has the form:
\begin{equation}
h=P_1(l_1,\ldots,l_n)\otimes T_1+\ldots +P_N(l_1,\ldots, l_n)\otimes T_N
\label{3}\end{equation}
 where $P_1,\ldots ,P_N$ are polynomial expressions 
of infinitesimal generators $l_i=\delta(e_i)$; $\{e_i\}$ 
is a basis in $L$; and  $T_1,...,T_N$ are Hermitian operators acting in 
$H_B$, $T_{\alpha}\in H_B$.
\end{description}

Systems with quantum  dynamical symmetry can  
 be explored by generalized methods known  from 
 the theory of systems possessing a  classical dynamical
 symmetry, e.g. by the method  of quantum 
coherent states \cite{Demichev}.
Let us observe  that 
the terms can be reorganized in such a way
 that the Hamiltonian has the familiar form: $
h=h_S+h_B+h_I$,
where $h_S$ is the  Hamiltonian of such system
(sum  of terms with $T_i=\mbox{\rm id}_{H_B}$), 
$h_B$ is the Hamiltonian of the environment 
(sum of the terms with constant parts of 
$\delta(P_i)$ which are  $\mbox{\rm id}_V$)
and $h_I$ is the interaction Hamiltonian. 

Let $v:V\to V\otimes {\cal A}$ 
be an arbitrary representation of 
$G$ in the finite-dimensional vector space $V$, 
and let $\delta:L\to \mbox{\rm End}(V)$
be the associated representation of $L$.
To further simplify the considerations, 
we shall consider the case in which the 
quantum group is `connected' in the sense that 
$ker(d)=C\hspace{-2.8mm}I 1_{\cal A}$.

Then for every vector $u\in V$, $v(u)=u\otimes 1_{\cal A}$
is equivalent to
 $\forall x\in L: \delta(x)u=0$.

Let us now assume that the calculus $\Gamma$ is also bicovariant. 
This enables us to introduce the quantum universal enveloping algebra $U(L)$, and to discuss the 
representations of $U(L)$ associated with
the representations of $G$. Let us 
introduce the map $\chi :U(L)\to C\hspace{-2.8mm}I$, 
with the properties $\chi(L)=0$, $\chi(1)=1$, and 
we extend it to $U(L)$ by multiplicativity. 
The representation $\delta$ uniquely (as in 
the standard theory) extends from $L$ to $U(L)$. The above two 
conditions are  equivalent to $\delta(q)u=\chi(q)u, \hspace{5mm}\forall q\in U(L)$

Vectors satisfying  any of the above 
conditions are called {\it v-invariant}. 
 Having such $v$-invariant vectors and 
an open system with quantum   dynamical
 symmetry 
one can prove:

{\bf Theorem 1} \\
The unitary evolution  described by the
 Hamiltonian $h$ of the  form 
(\ref{3}) preserves the $v$-invariance of the   
vectors and associated states of the 
system,  even when  all  other states 
of the system are 
corrupted due to decoherence.\\
{\bf Proof:}
Let us take  as an initial vector 
$u\otimes \zeta\in V\otimes 
H_B$, where $u$ is $v$-invariant 
in the sense defined above.
 Then the unitary evolution 
defined by: $U(t)=\exp(-\frac{i}{\hbar}ht)$ 
gives:
\begin{eqnarray}
& &\exp(-\frac{i}{\hbar}ht)(u\otimes \zeta)=u\otimes 
\exp(-\frac{i}{\hbar}h_{\scriptsize\rm eff}t)\zeta\\
&\mbox{where}&h_{\scriptsize\rm eff}=\chi(P_1)T_1+\ldots +\chi(P_N)T_N\hspace{2cm}\Box
\end{eqnarray}

Now we can easily  generalize 
Theorems 1 and 2 given in \cite{Rasetti}.
We follow the notation of \cite{Rasetti}.
$\rho_S\in \mbox{\rm End}(V)$ and $\rho_B\in \mbox{\rm End}(H_B)$
 are states of the system and the environment 
 (bath), respectively. If the overall 
system is initially in the state 
$\rho(0)=\rho_S\otimes \rho_B$, 
then $\rho(t)=U(t)\rho(0)U(t)^+$, 
so that the evolution is unitary.
 The induced evolution on $V$ 
is, similarly to that   in  
\cite{Rasetti}, given by 
$L_t^{\rho_B}:\rho_S\to tr^B \rho(t)$,
 where $tr^B$ is the trace over $H_B$. 
Then the following theorem can be proved:

{\bf Theorem 2} \\
Let ${\cal M}_N$ be the manifold of 
states built over the space  of vectors 
invariant under the 
representation $v$, and 
$\rho_S\in{\cal M}_N$.
Then for any initial bath state $\rho_B$ 
the induced evolution on $V$ is trivial, 
$L_t^{\rho_B}(\rho(t))=\rho$, $\forall t>0$.

Theorem 1 allows us to reduce the proof of
Theorem 2 to the proof of the first
theorem of \cite{Rasetti}. 
 
The invariant vectors are generalizations 
of the singlet states  in \cite{Rasetti} as the states of the 
quantum register which  are not corrupted  by the interaction with the environment.

Before we present simple examples 
illustrating the general theory  
and explicitly demonstrating  
``error-protected" states, let 
us discuss  the interesting question
  of the structure of the Hilbert
 space of the registers of the quantum computer. 
We will also  discuss the 
 physical implications.  The
 register usually consists 
of a number of copies of the 
same quantum system, often  
having two possible states, 
e.g. {\it  spin up} and  {\it spin  down} (qubit).

Dynamical symmetry acts in the Hilbert
 space that originates from the Hilbert
 space for an individual qubit being 
described as  a representation  space
 of our quantum  group $G$:
$v_i:V_i\to V_i\otimes{\cal A}$, $i=1,\ldots, n$. 
The Hilbert space is the tensor product of the 
representation spaces,  
$V=V_1\otimes V_2\otimes\ldots\otimes V_n$ 
on which  the tensor product of 
representations $v_i$, 
$v=v_1\times v_2\times \ldots \times v_n$ acts.
 Since with each of the representations 
$v_i$ is associated a representation 
 $\delta_i$ of the corresponding quantum 
universal enveloping  algebra, to the 
representation  $v$ corresponds  the 
representation $\delta$ of the  quantum
  universal enveloping algebra.  
One can prove \cite{Woronowicz3} 
the following relation (for $n=2$):
\begin{equation}
\delta(x)(\phi_1\otimes \phi_2)=\sum_\alpha\delta_1(x^\alpha)\phi_1
\otimes\phi_2^\alpha+
\phi_1\otimes\delta_2(x)\phi_2
\end{equation}
where $\tau(\phi_2\otimes x)=\sum_\alpha x^\alpha\otimes \phi_2^\alpha$ and 
$\tau : V_2\otimes L\to L\otimes V_2$
is the appropriate {\it flip-over} operator uniquely 
defined by the differential calculus. 

This formula differs from the corresponding formula for the classical
 case \cite{Louck}
of addition of  angular momenta in quantum mechanics ($\tau$
 in the classical case is just the standard 
transposition). Its diagrammatic representation 
\begin{equation}
\hbox{Fig. 1}
\end{equation}
and its  generalization to arbitrary $n$-fold coupling
\begin{equation}
\hbox{Fig 2}
\end{equation}
show that the qubits in the register are not treated on the same footing.  It
could be associated to some effects due to,  not taken into account
\cite{Rasetti}, linear  extension of the register, or to fluctuations  
of the fields due to nonideal structure of boundaries of the register and their 
influence, etc. It is possible then  to realize a system with weaker symmetry 
than the one presented in \cite{Rasetti}. It is known that similar
deviations from exact dynamical symmetry of Lie groups lead to 
better mass or energy formulas in  nuclear, particle, and
molecular physics, e.g. \cite{Iwao}, \cite{Gavrilik}.
Therefore, one can look  for possible candidates for
registers of quantum computers there.

In \cite{Rasetti} the  physically plausible conjecture was expressed that small 
deviations from ideal properties of the 
system  should lead to small errors  in the error-protected 
states. Actually, we have shown that there exist systems with 
special kind of deviation from the assumed symmetry, which
nevertheless still have error-protected states.

Let us examine some 
simple examples that illustrate our general ideas
 and  theorems. The first
 example of a quantum group 
presented systematically in 
the literature was the $S_{\mu}U(2)$ 
group  \cite{RIMS}, where the 
$C^*$-algebraic approach to 
quantum groups was used. 
In \cite{RIMS} not only the algebraic
 and functional analytic aspects were 
treated, but also the geometry of
 $S_{\mu}U(2)$, including the left-covariant
 three-dimensional calculus, and the 
bicovariant four-dimensional 
calculus (discussed also 
in detail in \cite{Stach}). 
Generalization of the results 
concerning this particular quantum 
group leads to the general theory of 
compact matrix quantum groups 
\cite{Woronowicz1,Woronowicz2}, to 
the definition of quantum spheres \cite{Podles1} and their geometry 
\cite{Podles2}, to deep generalization of the Tannaka-Krein duality 
\cite{Woronowicz3}, and also to  the theory of quantum principal 
bundles together with the corresponding gauge theory on quantum spaces, 
first formulated in \cite{Micho1} and then developed 
systematically in \cite{Micho2,Micho3} (see also \cite{Micho4,Micho5,Micho6}). 
Also in the $C^*$-algebraic 
framework,  quantum homogeneous 
bundles were defined and the example 
of such a bundle with quantum spheres
as fibers was given \cite{Hania}.
 Simultaneously, a different approach 
to  quantum groups was developed by 
Soviet \cite{Drinfel'd,Kulish} 
and  Japanese schools \cite{Jimbo},
 in which  quantum groups are treated 
from the point of view of deformations of the universal enveloping algebras. 
In the latter approach,  the utilization of  quantum groups  for studying 
completely solvable systems seems to be the main motivation for 
developing the theory.

In this paper, we  conceptually follow
the first of these approaches. We use the quantum group 
 $S_{\mu}U(2)$ in our examples. First, we recall
some basic facts about $S_{\mu}U(2)$
(the case $\mu \in [-1,1]$, $\mu =1$
corresponds to the classical $SU(2)$ group).
This quantum group is based on 
a $^*$-algebra $\cal A$
generated by elements $\alpha$, $\alpha^*$, 
$\gamma$, $\gamma^*$ satisfying the 
following relations:
\begin{equation}
\alpha\alpha^* + \mu^2\gamma^*\gamma=1_{\cal A},\,\,\,
\alpha^*\alpha + \gamma^*\gamma=
1_{\cal A},\,\,\,
\gamma^*\gamma=\gamma\gamma^*,\,\,\,
\alpha\gamma=\mu\gamma\alpha,\,\,\,
\alpha\gamma^*=\mu\gamma^*\alpha.
\end{equation}
The comultiplication, counit, 
and the antipode are defined on the generators of the algebra by:\\
i) comultiplication:\\
$\Phi(\alpha)=\alpha\otimes\alpha-\mu\gamma^*\otimes\gamma,\quad
\Phi(\gamma)=\gamma\otimes\alpha+\alpha^*\otimes\gamma$\\
with $\Phi(\alpha^*),\Phi(\gamma^*)$ 
fixed by the property $\Phi^*=(^*\otimes^*)\Phi$.\\
ii) counit:
$e(\alpha)=1\quad e(\gamma)=0$\\
iii) antipode:
$\kappa(\alpha)=\alpha^*,\quad
\kappa(\alpha^*)=\alpha, \quad
\kappa(\gamma)=-\mu\gamma,\quad
\kappa(\gamma^*)=(-1/\mu) \gamma^*$

From the point of 
view of our examples,  the theory 
of representations of $S_{\mu}U(2)$ is very interesting. 
The theory has many similarities to 
the theory of representations of 
$SU(2)$. Since the representations 
(also in the case of a general 
quantum group $G$) are linear
 maps $v:V\to V\otimes {\cal A}$, 
we can introduce a matrix 
representation in a given
 basis $\{e_i\}$ in $V$, by 
$v(e_i)=\sum_{j} e_{j} \otimes v_{ji}$.
 In the case in which the basis is 
orthonormal and $v$ is unitary, the
matrix of the representation $v$ is 
also unitary (in the extended sense 
in which the Hermitian conjugate of a 
matrix is the transposition of the matrix 
created by  $^*$-conjugation of its 
elements). As a basic example of such 
representation, consider the fundamental
 representation of $S_{\mu}U(2)$, 
defined in an orthonormal basis by the matrix:
\begin{equation}
u_{ij}=\pmatrix{\alpha&-\mu\gamma^*\cr\gamma&\alpha^*\cr}.\label{**}
\end{equation}
The reader can easily check using the 
 defining relations for $S_{\mu}U(2)$ 
that the matrix is unitary:$
u^*u=uu^*=1,$
where  the unit on the right side 
is actually the tensor product
of the unit in the $S_{\mu}U(2)$
with the unit $2\times2$ matrix.

In our  considerations 
it is essential
that the irreducible unitary representations 
of $S_{\mu}U(2)$ are associated with 
half-integers, like the 
representations of $SU(2)$. 
The fundamental 
representation introduced  by (\ref{**}) corresponds to $j={1\over2}$.
 The Clebsch-Gordan decompositions of
 tensor products of the representations
 of the $S_{\mu}U(2)$ into irreducible 
representations are similar 
(concerning the multiplicities of the appearence of irreducible
components in the products of representations) to the classical case:
\begin{equation}
u^{\otimes n}=\bigoplus_{j \in{J}}n_ju_j
\end{equation}
In particular, the decomposition 
of the second tensor power of the 
fundamental representation is 
$u_{\small 1/2}^{\otimes2}=u_0 \oplus u_1$,
 where $u_0$ and $u_1$ are the 
$1$-dimensional and the
 $3$-dimensional irreducible
 representations, respectively. 
One can describe these representations
 more explicitly after introducing 
an orthonormal basis in the 
representation space 
$V={C \hspace{-2.8mm}I} \hspace{0.03mm} ^2$ of $u_{\small 1/2}$, which will be denoted
 $\vert+\rangle$, 
$\vert-\rangle$ for the purpose of being familiar to 
 physicists. The tensor product $u_{\small 1/2} ^{\otimes2}$
 is realized in ${V \otimes V }\simeq {{C\hspace{-2.8mm}I} 
 \hspace{0.1mm} ^4}$, and the orthonormal basis in this space is $\vert+
\rangle\otimes\vert+\rangle$, $\vert+\rangle\otimes\vert-\rangle$, 
$\vert-\rangle\otimes\vert+\rangle$, $\vert-\rangle\otimes\vert-\rangle$. 
It is an easy exercise to show that the invariant
 subspaces of $u_{\small 1/2}^{\otimes2}$ are spanned by:
\begin{eqnarray}
& &{1 \over {\sqrt {1+{\mu}^2}}}(\vert +\rangle\otimes\vert -\rangle 
-\mu\vert -\rangle\otimes\vert +\rangle)
\label{singlet}\\\vert +\rangle\otimes\vert +\rangle,& &
{\mu \over {\sqrt {1+{\mu}^2}}} (\vert +\rangle\otimes\vert
 -\rangle +{1 \over {\mu}} \vert -\rangle\otimes\vert +\rangle ),\hspace{7mm}
\vert -\rangle\otimes\vert -\rangle\label{triplet}\end{eqnarray}

(\ref{singlet}) generalizes the singlet state, and 
(\ref{triplet}) generalizes the  triplet state. 
In analogy to the classical case, 
the even tensor powers of the fundamental representation 
decompose into irreducible 
representations in such a 
way that the one-dimensional representation appears a number 
of times, and the number is identical to the classical 
case. These singlets are  preserved by the dynamics.

{\bf Examples:}\\
1) In the first example we 
treat a system which is as
 close as possible to the 
one considered in \cite{Rasetti}.
 Namely, as a model of the
 environment (bath) we 
consider a system of harmonic 
oscillators, described by the
 Hamiltonian 
$h_B = \sum_{k} \omega_{k}b_{k}^{\dagger}b_{k}$,
acting in the Hilbert space $H_B$,
$h_B \in {\rm End}(H_B)$. 
The register consists in this simplest case of 
two qubits. In contrast to the case considered
by Zanardi and Rasetti \cite{Rasetti}, the system consisting of the 
register and the bath has the dynamical symmetry 
not of the $SU(2)$ group but of the $S_{\mu}U(2)$ 
quantum group. As previously  mentioned, 
in the quantum group
context it is necessary to choose a  
differential calculus, prior to establishing the notion of
the dynamical symmetry associated with  a given quantum group. 
The closest calculus to the classical case seems to be  the $3D$ left-covariant 
calculus \cite{RIMS}. In other words, the 
quantum Lie algebra $L$ is $3$-dimensional. 
Let us denote by $K_i$ the operators 
representing the basis vectors $l_i$, in an arbitrary 
representation of $L$ (here $i\in\{1,2,3\}$). 
The following recurrent formulas enable
us to compute explicitly the operators $K_i$, 
in the arbitrary tensor product of elementary
$2$-dimensional representations -- qubits (where $j\in\{1,2\}$):
\begin{eqnarray}
K_3(\psi\otimes\vert{+}\rangle)&=&\frac{1}{2}\psi
\otimes\vert{+}\rangle+\frac{1}{\mu^2}
K_3(\psi)\otimes\vert{+}\rangle\\
K_3(\psi\otimes\vert{-}\rangle)&=&\mu^2
K_3(\psi)\otimes\vert{-}\rangle-\frac{1}{2}\psi\otimes\vert{-}\rangle\\
K_j(\psi\otimes\vert{+}\rangle)&=&\frac{1}{2}\psi\otimes\vert{+}
\rangle+\frac{1}{\mu}
K_j(\psi)\otimes\vert{+}\rangle\\
K_j(\psi\otimes\vert{-}\rangle)&=&\mu K_j(\psi)\otimes\vert{-}\rangle
-\frac{1}{2}\psi\otimes\vert{-}\rangle,
\end{eqnarray}

In this case, the bath-register interaction Hamiltonian which
is the quantum group analog of the Hamiltonian used in \cite{Rasetti} 
$(h_I=\sum_kg_kS^+b_k+f_kS^-b_k^++h_kS^zb_k+H.c.)$              is:
$h_I=K_+T+K_{-}T^\dagger+K_3T',$
where $K_{\pm}=K_1\pm iK_2$, and $T,T'$ are operators acting in the bath
Hilbert state-space. 
The operators $T$ and $T'$ are obtained as  appropriate linear 
combinations of the creation and annihilation operators $(b_k,b_k^+)$ describing 
relevant elementary excitations of the bath.
The operators $K_j$  act in the 
$4$-dimensional 2-qubit space. $K_+,K_-$  coresspond  to classical $S^+,S_-$, $K_3$ 
corresponds to $S^z$. 
In other words, the Hamiltionial  is formally the same form as in \cite{Rasetti}. However the
`spin' operators are different as explained above. It is obvious that the 
singlet state of the register is error-protected in the 
sense discussed above.

2) In the second example 
the only change from 
 example 1) is the 
register consists of 
any even number of qubits,
instead of just two. The spin operators $K_j$  refer to the total
register system, and are calculated by applying the above  rules 
inductively. 

It is important to mention that the number of singlet states 
is  the same as in the 
classical $SU(2)$ case. This is a consequence of the  similarity between
the representation theories for quantum and classical $SU(2)$ groups.
The dimension of the singlet state space depends on the number of 
qubits in the way described 
in \cite{Rasetti}. All of these
states are clearly protected
from corruption due to decoherence. 

  Coupling of the qubits  to the
same environment gives
more  error-protected
states than coupling to independent 
environments  \cite{Chinese}. 
We made this assumption in our  examples, as did the authors of  \cite{Rasetti}. Nevertheless, our methods
are  general enough to deal
with the cases of coupling to
independent environments as well when 
the system has the  dynamical symmetry of
the type introduced in this paper.
After completion of the paper
 the authors found the paper 
\cite{Rasetti2} which extends 
 \cite{Rasetti},
 still in the context of the
classical group dynamical 
symmetry.

MD acknowledges the hospitality of LANL, USA. HM and RO acknowledge the  hospitality of
UNAM, Mexico. This research was 
partially supported by Investigation Project IN106879 of DGAPA/UNAM.

\end{document}